\documentclass[twocolumn]{aastex63}

\received{}
\revised{}
\accepted{}

\submitjournal{ApJ}

\shorttitle{Host galaxy of OJ~287}
\shortauthors{Nilsson et al.}

\begin{document}

\title{The host galaxy of OJ~287 revealed by optical and near-infrared imaging}

\correspondingauthor{Kari Nilsson}
\email{kani@utu.fi}

\author[0000-0002-1445-8683]{K. Nilsson}
\affiliation{Finnish Centre for Astronomy with ESO (FINCA)\\
Quantum, Vesilinnantie 5, FI-20014 Turun yliopisto, Finland}

\author{J. Kotilainen}
\affiliation{Finnish Centre for Astronomy with ESO (FINCA)\\
Quantum, Vesilinnantie 5, FI-20014 Turun yliopisto, Finland}
\affiliation{Department of Physics and Astronomy, Vesilinnantie 5, FI-20014 Turun yliopisto, Finland}

\author{M. Valtonen}
\affiliation{Finnish Centre for Astronomy with ESO (FINCA)\\
Quantum, Vesilinnantie 5, FI-20014 Turun yliopisto, Finland}

\author{J. L. Gomez}
\affiliation{Instituto de Astrofísica de Andalucía (CSIC), Glorieta de la Astronomía s/n, Granada 18008, Spain}

\author{A. J. Castro-Tirado}
\affiliation{Instituto de Astrof\'isica de Andaluc\'ia (IAA-CSIC), P.O. Box 03004, E-18080, Granada, Spain}

\author{M. Drozdz}
\affiliation{Mt. Suhora Observatory, Pedagogical University, ul. Podchorazych 2, 30-044 Krakow, Poland}

\author{A. Gopakumar}
\affiliation{Department of Astronomy and Astrophysics, Tata Institute of Fundamental Research, Mumbai 400005, India}

\author{S. Jeong}
\affiliation{Department of Physics, SungKyunKwan University, Suwon 16419, Korea}

\author{M. Kidger}
\affiliation{Herschel Science Centre, ESAC, European Space Agency, E-28691 Villanueva de la Cañada, Madrid, Spain}

\author{S. Komossa}
\affiliation{Max Planck Institut fuer Radioastronomie, Auf dem Huegel 69,53121 Bonn, Germany}

\author{S. Mathur}
\affiliation{Department of Astronomy and Center for Cosmology and AstroParticle Physics, The Ohio State University, 140 West 18th Avenue,Columbus, OH 43210, USA}

\author{I. H. Park}
\affiliation{Department of Physics, SungKyunKwan University, Suwon 16419, Korea}

\author{D.E. Reichart}
\affiliation{University of North Carolina at Chapel Hill, Chapel Hill, North Carolina NC 27599, USA}

\author{S. Zola}
\affiliation{Astronomical Observatory of the Jagiellonian University, ul. Orla 171, PL-30-244 Krakow, Poland}
\affiliation{Mt. Suhora Observatory, Pedagogical University, ul. Podchorazych 2, 30-044 Krakow, Poland}

\begin{abstract}

The BL Lacertae object OJ 287 (z = 0.306) has unique double-peaked
optical outbursts every $\sim$12 years, and it presents one of the
best cases for a small-separation binary supermassive black hole
(SMBH) system, with an extremely massive primary $\log (M_{\rm
  BH}/M_{\odot}) \sim$ 10.3. However, the host galaxy is unresolved or
only marginally detected in all optical studies so far, indicating a
large deviation from the bulge mass - SMBH mass relation. We have
obtained deep, high spatial resolution i-band and K-band images of
OJ~287 when the target was in a low state, which enable us to detect
the host galaxy.  We find the broad-band photometry of the host to be
consistent with an early type galaxy with $M_{\rm R} = -22.5$ and
$M_{\rm K} = -25.2$, placing it in the middle of the host galaxy
luminosity distribution of BL Lacertae objects. The central
supermassive black hole is clearly overmassive for a host galaxy of
that luminosity, but not unprecedented, given some recent findings of
other ``overmassive'' black holes in nearby galaxies.

\end{abstract}


\section{Introduction} \label{sec:intro}

Spheroidal galaxies and galactic bulges harbor central supermassive
black holes (SMBH), and there is a fundamental tight relation between
the SMBH mass and the bulge luminosity (mass), both in quiescent and
active galaxies
\cite{2009ApJ...698..198G,2011Natur.469..377K,2014MNRAS.445.1261S}. Methods
to measure the central BH mass include stellar and gas kinematics
\citep[e.g.][]{2009ApJ...700.1690G}, megamasers
\citep[e.g.][]{2011ApJ...727...20K}, reverberation mapping
\citep[e.g.][]{2000ApJ...533..631K} and emission line widths in
quasars \citep[e.g.][]{2012AdAst2012E..18D,2014MNRAS.445.1261S}. Given
that galaxy mergers are thought to be common, especially at high
redshift \citep[e.g.][]{2008MNRAS.391..481S}, it is reasonable to
expect that double, even multiple BH systems form in galactic nuclei.

The BL Lacertae object OJ 287 (z = 0.306) presents one of the best
cases for a small-separation SMBH binary system
\citep[e.g.][]{2008Natur.452..851V}, another examples being
SDSS~J120136.02+300305.5
\citep{2012A&A...541A.106S,2014ApJ...786..103L} and possibly GSN 069
\citep{2019Natur.573..381M}.  Its historical light curve is unique
among active galaxies, due to the prominent recurring double-peaked
optical outbursts every $\sim$12 years. Each outburst consists of two
rapid brightenings, separated by $\sim$1 year and lasting $\sim$6
weeks \citep{1996A&A...315L..13S,2013A&A...557A..28V}. A notable
feature of the outbursts is that they are not strictly periodic, nor
is the separation of the two peaks constant from one outburst to the
next.  This behavior is attributed to strong precession of the
secondary's orbit in the model. We also note that the redshift of
OJ~287 is secure, based on the detection of several clearly
identifiable emission lines \citep[e.g.][]{2010A&A...516A..60N}.

The binary BH model of OJ~287 is very successful in reproducing the
previous outbursts, and it makes specific predictions about future
outbursts. It is extremely unlikely that a single BH model would
produce such good light curve predictions as the binary BH model has
done \citep{2011ApJ...729...33V}. In the binary BH model, the
best-fitting orbital parameters correspond to mass of the primary BH:
M = 1.84 $\times$ 10$^{10}$ $M_{\odot}$, and mass of the secondary BH:
m = 1.30 $\times$ 10$^8$ $M_{\odot}$. Notably, the primary SMBH is at
the high mass end of the quasar BH mass function
\citep[e.g][]{2008ApJ...674L...1V,2012MNRAS.420..732P}.

If OJ~287 contains an extremely massive primary BH, and if it follows
the tight BH - bulge relation of active galaxies \citep[e.g.][]
{2012AdAst2012E...3D}, its host galaxy is also expected to be
extremely luminous and massive. Puzzlingly, there have only been
marginal or non-detections of the host galaxy of OJ 287 based on
optical imaging
\citep[][]{1996ApJ...464L..47B,1996ApJS..103..109W,1997ApJ...484L.113Y,1999A&A...352L..11H,
 2000ApJ...532..816U,2002A&A...381..810P,2003A&A...400...95N}.
The reported host luminosities in these studies are $-22.0 <$ M$_{\rm
  R}$ $<$ -23.5, much fainter than expected from the BH–bulge
relation, M$_{\rm R}$ $\sim$ -26). Converting the host galaxy absolute
magnitudes into stellar mass by adopting the M/L of a single stellar
population originated at z(burst) = 5 and passively evolving down to z
= 0 \citep{2010MNRAS.402.2453D}, these luminosities translate into
host galaxy mass in the range of $\log (M_{\rm host}/M_{\odot})$ =
11.4--11.9, much smaller than the expected host galaxy mass based on
the BH–bulge relation, $\log (M_{\rm host}/M_{\odot})$ $\sim$
13. Conversely, based on the marginal host galaxy detections in the
optical, if OJ~287 follows the BH–bulge relation, the primary BH mass
is only $\log (M_{\rm BH}/ M_{\odot})$ = 8 -- 9. Interestingly, this
value is within the range of typical BH masses for BL Lac objects
\citep{2003ApJ...595..624F,2011MNRAS.413..805P}, but much smaller than
required in the binary SMBH model.

There is a reported near-infrared (NIR) detection of the host galaxy
of by \cite{1998MNRAS.295..799W}, who found for the host galaxy to be
extremely bright, $M_{\rm K}$ = -28.9 $\pm$ 0.6. Transforming this to
our cosmology we obtain $M_{\rm K}$ = -28.5 $\pm$ 0.6. With this
luminosity, OJ~287 would fall exactly on the host galaxy luminosity -
SMBH mass relation.  However, the implied color of the host galaxy, R
- K = 6.1 , would be much redder than in a typical elliptical galaxy
(R - K = 2.7).  We note that the data were taken with one of the first
generation NIR imagers (IRCAM 62 $\times$ 58 px) and that they could
not resolve the host galaxy in the R-, J- or H-band with observations
under similar seeing conditions and with similar exposure times as in
the K-band. Furthermore, the relatively poor spatial resolution
(0.3-0.6 arcsec pixels and 0.9-1.3 arcsec seeing) and poor definition
of the PSF severely limit their accuracy.

In this paper we present new observations, which enable us to detect
the host galaxy of OJ~287 both in the optical and in Near-IR. This
enables us to constrain the host galaxy luminosity further and to shed
some light into some conflicting results obtained so far. In all
calculations w e use the cosmology $H_0 = 70$ km s$^{-1}$ Mpc$^{-1}$,
$\Omega_{\Lambda} = 0.67$ and $\Omega_{\rm M}$ = 0.33.

\section{Observations and data reduction}\label{sec:obs}

OJ~287 went into a deep minimum in November 2017, fading to R $\sim$
15.8 (S. Zola, priv. comm.), which is about 1 mag fainter than on
average between the flares. We exploited the favorable AGN/host
brightness ratio by obtaining a deep SDSS i-band image at the Gran
Telescopio Canarias (GTC) on the night of Dec 2, 2017 when OJ~287 was
already coming up from the minimum, but still about R $\sim$ 15.2. We
obtained 77 exposures of 1.5 seconds with the OSIRIS instrument with a
pixel scale of 0.254 arcsec/pix, totalling in 115.5 seconds of
exposure. The images were bias subtracted and flat-fielded with
twilight flats and registered using stars in the field of view. The
images were then co-added and deviations from an uniform background
were removed by fitting a low-order polynomial to the background
pixels and subtracted.  The co-added image has a FWHM of 0.78
arcsec. Calibration of this image was obtained through stars 15 and 16
in \cite{2001AJ....122.2055G} using SDSS DR14 published
magnitudes. The zero points derived from these two stars differed by
0.01 mag.

K$_{\rm s}$-band observations were obtained at the 2.5 m Nordic
Optical Telescope (NOT), La Palma, during the night of December 18,
2012. At this epoch OJ~287 was also at relatively low state at R
$\sim$ 15.4.  We used the 1024 $\times$ 1024 pixel NOTCam detector in
imaging mode with a pixel scale of 0.235 arcsec px$^{-1}$, giving a
field of view of $\sim$4 $\times$ 4 arcmin$^2$. The seeing during the
observations was $\sim$0.8 arcsec. The images were acquired by
dithering the target across the array in a random grid within a box of
$\sim$20 arcsec, and taking a 14.4 sec exposure at each position.
Individual exposures were then co-added to produce the final frame. A
total of 237 exposures of 14.4 sec were acquired, which provided a
total exposure time of 3412 sec.

Data reduction was performed using IRAF\footnote{IRAF is distributed
  by the National Optical Astronomy Observatories, which are operated
  by the Association of Universities for Research in Astronomy, Inc.,
  under cooperative agreement with the National Science Foundation.}
and followed the procedure described in \cite{2007ApJ...660.1039K} and
\cite{2007A&A...462..525H}. In each image, bad pixels were corrected
for using a mask made from the ratio of two sky flats with different
illumination levels. Sky subtraction was performed for each science
image using a median averaged frame of all the other temporally close
frames in a grid of eight exposures. Flat fielding was performed using
normalized median averaged twilight sky frames with different
illumination levels. Finally, images were aligned to sub-pixel
accuracy using field stars as reference points and combined after
removing spurious pixel values to obtain the final reduced co-added
image.

For photometric calibration, we used stars 16 and 17 in
\cite{2001AJ....122.2055G}, imaged simultaneously with the science
observations. The zero points derived from these two stats deviate
0.03 mag from each other, which is $\sim$1$\sigma$ difference
considering the error bars in \cite{2001AJ....122.2055G}.  Possible
small systematic error may remain due to the deviation of OJ~287
spectrum from a stellar spectrum and from the fact that
\cite{2001AJ....122.2055G} list K-band magnitudes and we used the
Ks-band filter.

\section{Analysis}\label{sec:analysis}

We analyzed the images by fitting two-dimensional surface brightness
models to the observed light distribution of OJ~287. The models
consisted of two components, an unresolved AGN nucleus and a host
galaxy. The former was described by three parameters, x-y position and
the flux in mJy and the latter by four parameters, x-y position, flux
and effective radius. Given that the host galaxy is very weak and
difficult to detect, we abandoned attempts to characterize it in more
detail. Therefore, the host was assumed to be a early-type galaxy with
S\'{e}rsic index equal to 4 and zero ellipticity. These assumptions
are quite reasonable since BL Lac nuclei are almost exclusively found
in giant ellipticals with S\'{e}rsic indices concentrated around $n =
4$ \citep[e.g][]{2000ApJ...542..731F,2003A&A...400...95N}. On the
other hand, given the special nature of OJ~287, it may not be hosted
by an ``ordinary'' galaxy. We will discuss this point later.

The faintness of the host galaxy and the inevitable errors in the PSF
model make it very difficult to reliably establish any offset $\leq$
FWHM between the host and the AGN nucleus. We thus assume that the AGN
nucleus is centered on the host galaxy.  The total number of
parameters is therefore 5: AGN/host $x$, AGN/host $y$, AGN flux, host
galaxy flux and host galaxy effective radius.

Prior to fitting, we performed three critical steps: masking, sky
background subtraction and PSF construction. Firstly, we masked out
any pixels with visible contribution from targets other than OJ~287.
Then we determined the sky background by measuring the average pixel
brightness in $\sim$10 rectangular clean areas around OJ~287, and then
computing their average. We then checked the obtained sky value by
plotting line and column cuts near OJ~287, looking also for any slope
in the background that might have remained from previous steps.

\begin{table*}[htb]
  \caption{\label{resulttable}Results of the model fitting}
  \begin{tabular}{llllll}
    Band & PSF star & PSF star mag & m$_{\rm AGN}$ & m$_{\rm host}$ & r$_e$\\
         &     & (mag)   & (mag)       & (mag)        & (arcsec)\\
    \hline
    i      & 15  &  15.49 & 14.83 $\pm$ 0.01 & 18.71 $\pm$ 0.09 & 1.3 $\pm$ 0.2 \\
           & 16  &  15.67 & 14.92 $\pm$ 0.01 & 18.30 $\pm$ 0.05 & 1.4 $\pm$ 0.2 \\
           & 17  &  15.60 & 14.82 $\pm$ 0.01 & 18.73 $\pm$ 0.10 & 0.9 $\pm$ 0.1 \\
K$_{\rm s}$ &  16  &  13.13 & 11.96 $\pm$ 0.01 & 15.27 $\pm$ 0.08 & 1.5 $\pm$ 0.3 \\
           &  17  &  14.02 & 12.19 $\pm$ 0.02 & 14.25 $\pm$ 0.10 & 0.5 $\pm$ 0.1 \\    
    \hline
  \end{tabular}
\end{table*}

Obtaining a good PSF was complicated by the fact that OJ~287 was
$\sim$2 times brighter than any star in the field of view in both
bands. This means that using a stellar image directly as the PSF model
does not work very well; the noise in the PSF image gets amplified
when the PSF is scaled to the brightness of OJ~287 and both the fit
and the residuals become very noisy. Analytical models are noise-free
and therefore don't have this problem. However, stellar images give a
more accurate PSF models than analytical models, so they are to
be preferred, if possible. 

Since the PSF is often a function of the position on the detector, we
constructed PSF models of several stars in the field of view. More
specifically we used stars 15, 16 and 17 in the i-band and 16 and 17
in the K-band. We first fitted an analytical model consisting of two
Moffat profiles to the stars. Subtracting this model revealed some
asymmetric features plus the diffraction pattern from the secondary
mirror support. These residuals were filtered using a wavelet-based
multiscale filtering technique \citep{1998ASPC..145..449M}, which
recovers any significant signal in the residual image with high
fidelity.  This filtered image was then summed to the analytical
model. In this way we obtain a low-noise PSF image, which accurately
preserves the deviations from perfect symmetry.

The model fits were made with an affine-invariant Monte Carlo Markov
Chain (MCMC) ensemble sampler \citep{2010CAMCS...5...65G}.  The
posteriori probability distribution was first sampled using this
sampler with multiple walkers. The number of walkers was equal to
three times the number of free parameters. We used Gaussian priors
which were not very restrictive, but nevertheless rejected
unreasonable parameter values, like for example parameters which would
place the model several pixels away from the observed position. The
host galaxy priors we especially carefully constructed so that the
priors would not dominate the fit over data.  Typically 5000-10000
iterations we made during each fit. After the sampler had finished, we
checked that the success rate for most of the walkers was close to
50\%. Then the marginalized histograms for each parameter were
extracted and the best-fit parameter values and their errors were
derived by fitting a Gaussian to these histograms.

The following noise model was used to compute the likelihood:
A pixel with a brightness of $N$ ADUs had an associated uncertainty
\begin{equation}
  \sigma = \sqrt{ \sigma_N^2 + \sigma_{\rm sky}^2 + \sigma_{\rm PSF}},
\end{equation}
where
\begin{equation}
  \sigma_N = \frac{\sqrt{(gN)^2+R^2}}{gs}
\end{equation}
is the photon and readout noise term, where $g$ is the effective
gain, $R$ the effective readout noise and $s$ is a constant ($\sim$
1.2 in this case) taking into account the smoothing caused by
interpolation when the images were registered. The term
\begin{equation}
  \sigma_{\rm sky} = \frac{\sigma_{\rm bg}}{c},
\end{equation}
where $\sigma_{\rm bg}$ is the background rms and $c$ is a constant
parametrizing our estimate of the background accuracy. Here we use $c
= 10$.  The last term $\sigma_{\rm PSF}$ represents the PSF
uncertainty and it is obtained from
\begin{equation}
  \sigma_{\rm PSF} = \sqrt{ \sigma_{sky}^2 + (d \cdot N)^2  },
\end{equation}
where $d$ is a constant, whose value was determined during the fit
process (see below). This PSF error model, while simplified, captures
one crucial aspect, namely that that far from the PSF center the sky
uncertainty dominates the PSF error, whereas close to the PSF core,
other factors dominate.

The fit proceeded as follows: First the value of $d$ was determined in
both bands by fitting a pure point source model to a star in the
field. We used stars 15 and 17 in the SDSS i-band and star 17 in the
K$_{\rm s}$ band. This fit had four free parameters, $x$, $y$, flux
and $d$.  This fit yielded $d = 0.05$ and 0.06 for the SDSS-i and
K$_{\rm s}$-bands, respectively. Then we fitted OJ~287 with the 5
parameter model keeping the $d$ value constant.

\begin{figure*}
  \hspace*{5mm}  
  \includegraphics[width=0.47\textwidth]{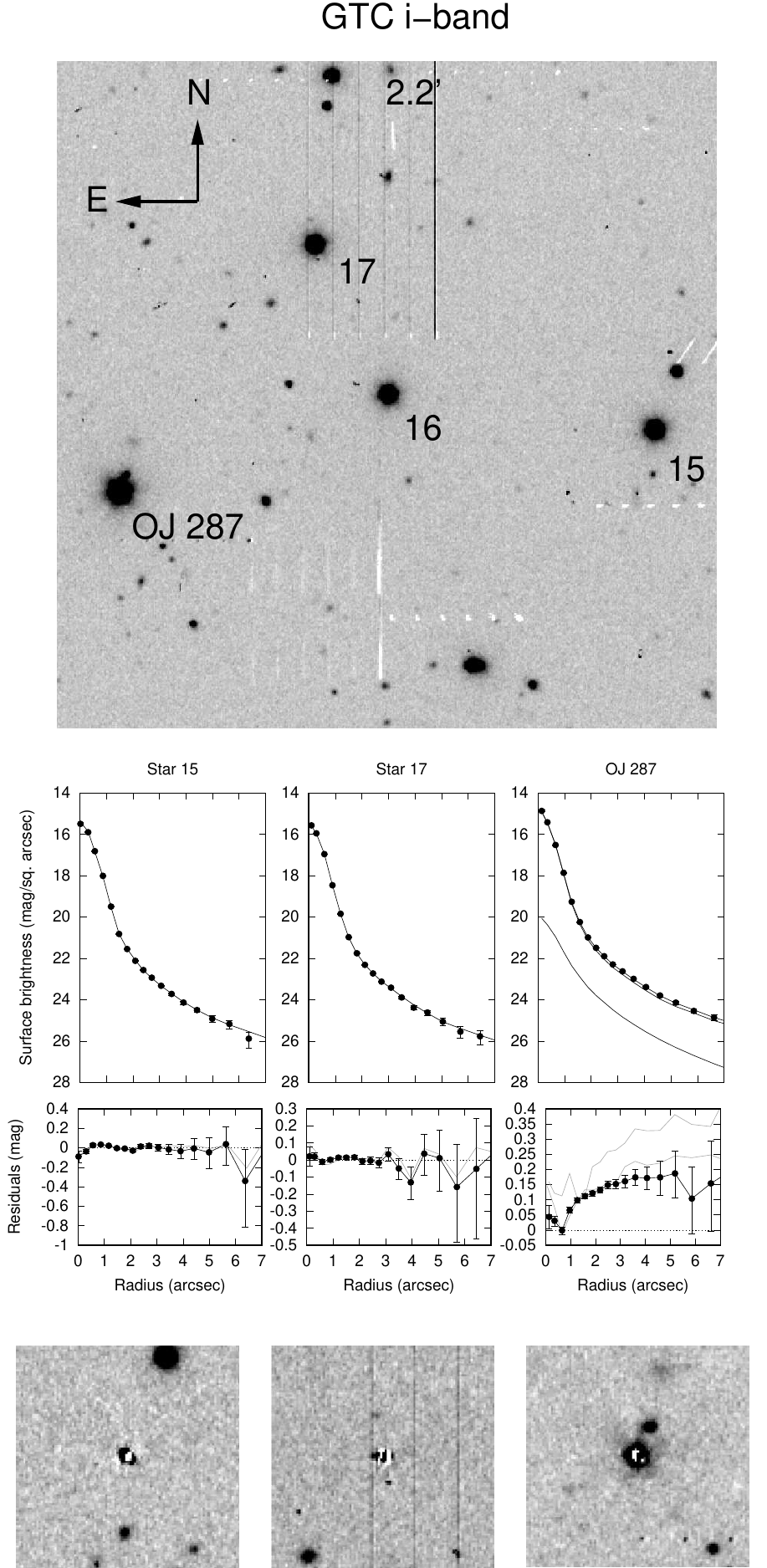}
  \hspace*{10mm}
\includegraphics[width=0.405\textwidth]{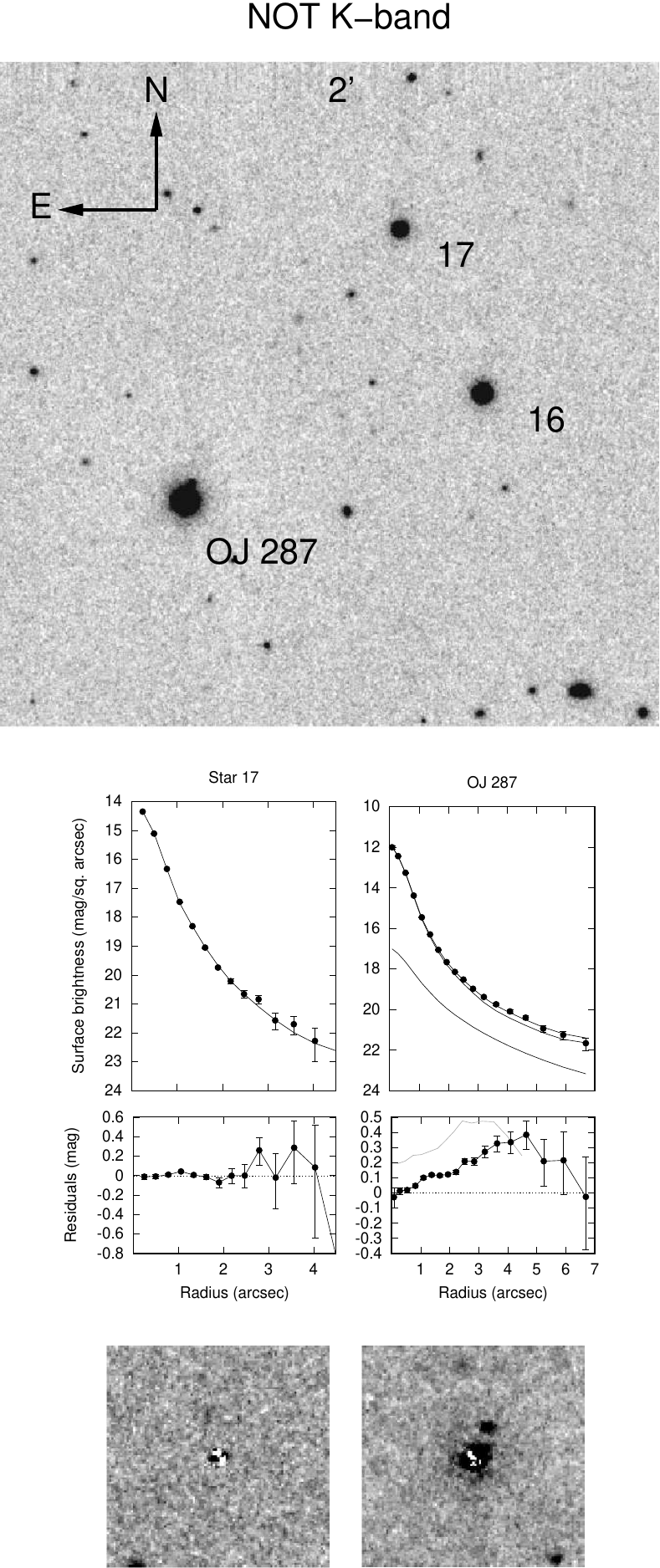}  
\caption{\label{profiles} Model fitting results for the GTC i-band
  (left) and the NOT K-band (right). {\em Upper panels}: Finding chart
  with targets discussed in the text identified.  {\em Middle two
    panels}: Radial surface brightness profiles of stars and OJ~287
  extracted from the images (solid symbols and the error bars). The
  solid lines show the best-fit model for each target using star 16 as
  the PSF (and thus the profile of this star is not plotted). In
  addition, the best-fit AGN component (dashed line) and host galaxy
  component (dot-dashed line) are shown for OJ~287. Below the profiles
  are the residuals after subtracting the scaled PSF. Black lines are
  using star16 as PSF, grey lines using stars 15 and 17. {\em Lower
    panels}: Residual images after subtracting the scaled PSF. The
  field size is 27$\times$27 arcsec in the i-band and 23$\times$23
  arcsec in the K-band. }
\end{figure*}

\section{Results}\label{sec:results}

The results of model fitting are summarized in Table \ref{resulttable}
and Fig. \ref{profiles}. There is a small but significant excess over
the PSF in both bands, visible in both the 1-d profiles and in the 2-d
residuals. This excess is best visible in the lower portion of
  the middle panel, which shows the residuals after subtracting the
  scaled PSF. Since the host galaxy is finite in size, we expect these
  residuals to tend towards zero at distances of a few effective
  radii, about 3-5 arcsec in case the of OJ~287. In the K-band the
  excess starts to turn down at $r \sim 4$ arcsec, but in the r-band
  there is no clear turning.

  However, none of the stars show any hint of an excess. Furthermore,
  far from the center, the error bars are dominated by the uncertainty
  of the sky level, which affects all points similarly. For instance,
  if the sky level is increased by 1 sigma, the residual plots would
  follow the lower ends of the error bars. Thus within one sigma sky
  uncertainty a turning down is possible in both bands. Thus we
  conclude that there is a excess of light around OJ~287 with respect
  to the stars in the vicinity, but this excess is rather weak and
  difficult to characterize.

Modeling the excess as a de Vaucouleurs profile, we obtain the values
in Table \ref{resulttable}. Two things can be read from this
table. Firstly, changing from one PSF star to another changes the host
galaxy properties more than indicated by the error bars derived from
the posteriori. This is especially true to the host galaxy magnitude
and it is a clear indication that the dominant source of error is the
PSF error.

Secondly, the host galaxy magnitude is less constrained in the
K-band. This is not due to the K-band observation having a less
favorable AGN/host flux ratio. Both images were taken with similar AGN
brightness and since the host galaxy is brighter in the K-band than in
the i-band, the core/host flux ratio is more favorable in the
K-band. The reason for worse performance in the K-band is most likely
due to the fact that the PSF stars are relatively fainter in the
K-band than in the i-band.  Thus when the PSF is scaled to the
brightness of OJ87, the inevitable errors in the PSF model are
amplified more in the K-band than in the i-band.

For the discussion below we convert the magnitudes to fluxes and take
the mean of the results in table \ref{resulttable} in both bands. The
error bars are set to the smallest number that keeps all results
within the error bars.  Thus in what follows we use $f_{\rm host}(i) =
(140 \pm 40) \mu$Jy and $f_{\rm host}(K) = (930 \pm 410) \mu$Jy.

\begin{figure}
  \includegraphics[width=0.5\textwidth]{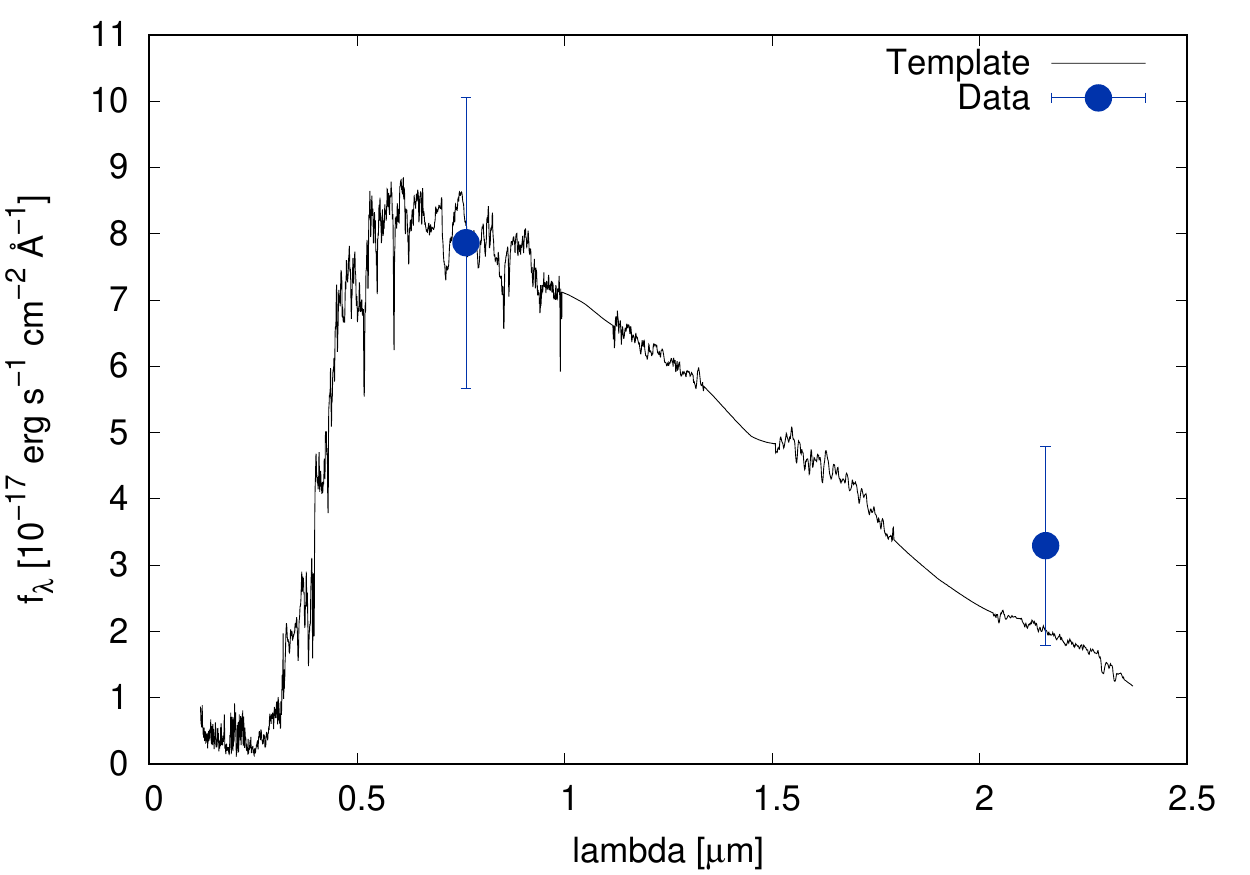}
\caption{\label{fluxplot} Broadband fluxes of OJ~287 host galaxy
    (filled symbols) with the rest-frame spectrum of an early-type
    galaxy from \cite{2001MNRAS.326..745M}. The scaling of the
    spectrum is obtained through a fit to the two data points.}
\end{figure}

In Fig. \ref{fluxplot} we show the rest-frame broadband flux of the
host galaxy as a function of wavelength. We also overplot an
early-type galaxy spectrum at zero redshift. To transfer the data to
the same redshift, we need to apply the K-correction $K(z)$ and
evolution correction $e(z)$, both of which depend on the filter and
galaxy type.  We assume here that the host galaxy is an early type
galaxy and obtain $K(z)$ from \cite{2010MNRAS.405.1409C}. The
K-correction has almost no dependence on galaxy type in the K-band
and also in the SDSS r-band, which we use as a proxy for the R band,
the dependence is mild, except for the starburst galaxies. There appears
to be very little star formation going on in
OJ~287. \cite{2010A&A...516A..60N} measured a narrow H$_{\alpha}$ line
flux of $4 \times 10^{-16}$ erg s$^{-1}$ cm$^{-2}$, which transforms
to a H$_\alpha$ line luminosity $L(H_{\alpha}) = 1.6 \times 10^{40}$
erg s$^{-1}$. Using the formula for the star forming rate SFR = $4.6
\times 10^{-42} L(H_{\alpha})$ M$_{\odot}$ yr$^{-1}$
\citep{2012MNRAS.420.1061T}, we obtain SFR = 0.07 M$_{\odot}$
yr$^{-1}$. This has to be considered as a generous upper limit since
the narrow H$_{\alpha}$ line could be produced mostly by the AGN
activity. We thus do not consider the starburst models in
\cite{2010MNRAS.405.1409C} and obtain $K(z)$ = 0.45, 0.25 and -0.5 for
the R, SDSS i and K-bands, respectively.  For the evolution
correction, we use $e(z)$ = 0.2 for both bands.  This value was
determined by running the Pegase 3 code \citep{2019A&A...623A.143F}
for a scenario with a single starburst 10 Gyr ago and passive
evolution thereafter.  The vertical scaling of the template in
Fig. \ref{fluxplot} was determined by integrating the template over
the respective bandpasses and finding the scaling factor that
minimizes the chi squared between our i and K-band observations and
the template fluxes.

\section{Discussion}\label{sec:discussion}

The observed fluxes are consistent with an early-type galaxy spectrum,
corrected for passive evolution from z=0.306
(Fig. \ref{fluxplot}). The scaled template corresponds to a galaxy
with $M_R = -22.5$, which is within the range of BL Lac host galaxy
luminosities $\langle M_R \rangle$ =-22.8 $\pm$ 0.5 found by
\cite{2005ApJ...635..173S}. The K-band luminosity of the scaled
template is $M_K = -25.2$ with (R - K) = 2.7.  For the effective radii
we find $r_e$ = 5 $\pm$ 2 kpc in the SDSS i-band and $r_e$ = 4 $\pm$ 3
kpc in the K$_{\rm s}$ band. both are consistent with sizes found for
BL Lac host galaxies, although the errors are so high that we can only
rule out a very large galaxy. Therefore, the host galaxy of OJ~287
does not look special in any particular way in comparison to BL Lac
host galaxies in general.

\begin{figure}
  \includegraphics[width=0.45\textwidth]{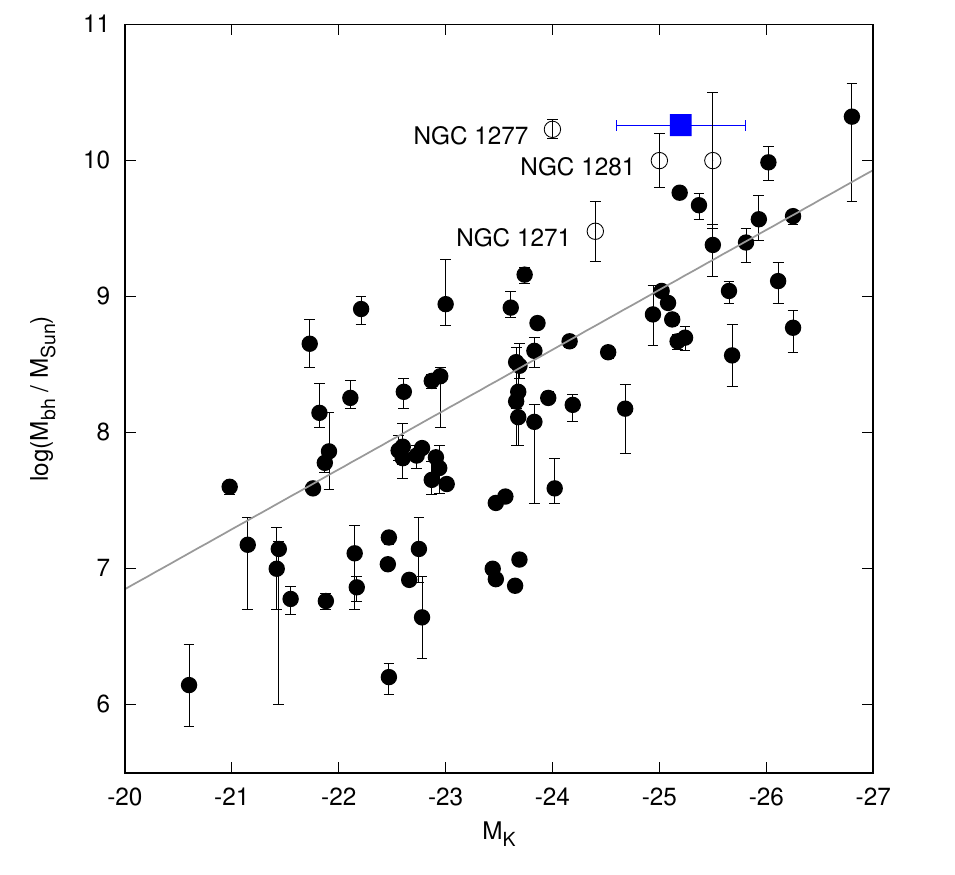}
  \caption{\label{mbhplot} Central supermassive black hole mass versus
    K-band luminosity of the host galaxy for the sample of nearby
    galaxies in \cite{2013ApJ...764..151G} (filled),
    some recently found massive SMBs (open)
    and OJ~287 (blue square). The gray line
    shows the best-fit K-band correlation from \cite{2013ApJ...764..151G}
    to the ``core-S\'ersic'' galaxies of their sample.
    The unlabeled open symbol denotes SDSS J151741.75-004217.6.
  }
\end{figure}

We next study how our results relate to the SMBH mass $1.84 \times
10^{10}$ M$_{\odot}$ indicated by the binary black hole model.  In
Figure \ref{mbhplot} we plot the host galaxy luminosity - SMBH mass
data from \cite{2013ApJ...764..151G} together with some recently found
very massive SMBHs and OJ~287, assuming for the latter the mass
indicated by the binary model and the luminosity of our scaled
template. The line in Fig. \ref{mbhplot} show the fit to the
``core-S\'ersic'' galaxies of the \cite{2013ApJ...764..151G} sample,
i.e. essentially to galaxies brighter than $M_{\rm K}$ = -23.5.  The
SMBH in OJ~287 is overmassive by 1.13 dex, roughly 2.6 times the
vertical rms scatter, 0.44 mag,in this plot, compared to SMBHs found
in luminous nearby bulges studied by \cite{2013ApJ...764..151G}.

The central SMBH in OJ~287 is thus at the high mass end of SMBHs with
a similar host, but not significantly deviating from the distribution. The
BH mass is basically a result of applying general relativity to the
timing of the flares and it is very accurately determined
\citep[e.g.][]{2008Natur.452..851V}.  Furthermore, such over-massive
SMBHs, challenging the co-evolution between SMBHs and their host
galaxy, have recently been found in an increasing number of galaxies,
e.g., NGC 4486B \citep{2011Natur.480..215M}, NGC1277
\citep{2012Natur.491..729V}, NGC1281 \citep{2016MNRAS.456..538Y} ,
NGC1332 \citep{2011MNRAS.410.1223R}, NGC 1271
\citep{2015ApJ...808..183W} and SDSS J151741.75-004217.6
\citep{2013MNRAS.434L..31L}. However, to our knowledge no such
detections have ever been made in supposedly spheroidal, massive host
galaxies of bright AGN.

According to \cite{2015ApJ...808...79F}, such galaxies form their
SMBHs early (z$\sim$2) and then experience lack of merger activity and
subsequent galaxy growth, preventing them from reaching the present-day
scaling relation. They are compact (R $<$ 2 kpc) and consist of purely
old stellar population (t $>$ 10 Gyr), and may be the local analogs
(relics) of z$\sim$2 massive galaxies
\citep[e.g.][]{2008ApJ...677L...5V,2007MNRAS.382..109T}.
Interestingly, these galaxies, including OJ 287, are at the expected
location of the M(BH)-Mbulge scaling relation at high redshift
\citep[e.g.][]{2010MNRAS.402.2453D,2012MNRAS.420..732P}.  Another way
to produce an undermassive host galaxy is via tidal stripping
\citep[see e.g.][]{2015ApJ...798...54G}, but this applies only to the
case of stellar mass loss due to tidal interaction with a more massive
galaxy, which is not the case for OJ 287.

\begin{acknowledgments}

Based on observations made with the Nordic Optical Telescope, operated
by the Nordic Optical Telescope Scientific Association at the
Observatorio del Roque de los Muchachos, La Palma, Spain, of the
Instituto de Astrofisica de Canarias. JK acknowledges financial
support from the Academy of Finland, grant 311438.  AJCT acknowledges
support from the Spanish Ministry Projects AYA2012-39727-C03-01 and
2015-71718R.  IHP acknowledges support from NRF
2018R1A2A1A05022685. SZ was supported by NCN grant
No. 2018/29/B/ST9/01793. The work is partly based on the observations
made with the Gran Telescopio Canarias (GTC), installed in the Spanish
Observatorio del Roque de los Muchachos of the Instituto de
Astrofisica de Canarias, in the island of La Palma.

\end{acknowledgments}

\bibliography{ojhost_v5}{}
\bibliographystyle{aasjournal}

\end{document}